\documentstyle[12pt]{article}
\begin{document} 
\thispagestyle{empty} 
\begin{flushright}
UA/NPPS-7-02\\
\end{flushright}
\begin{center}
{\large{\bf BOOTSTRAPING THE QCD CRITICAL POINT\\}}
\vspace{2cm} 
{\large N. G. Antoniou and A. S. Kapoyannis}\\ 
\smallskip 
{\it Department of Physics, University of Athens, 15771 Athens, Greece}\\
\vspace{1cm}

\end{center}
\vspace{0.5cm}
\begin{abstract}
It is shown that hadronic matter formed at high temperatures, according
to the prescription of the statistical bootstrap principle, develops a
critical point at nonzero baryon chemical potential.
The location of the critical point in the phase diagram, however, depends on
the detailed knowledge of the partition function of the deconfined phase,
near the critical line.
In a simplified version of the quark-gluon partition function, the resulting
location of the critical point is compared with the solutions of other
approaches and in particular with the results of lattice QCD. The
proximity of our solution to the freeze-out area in heavy-ion experiments is
also discussed.
\end{abstract}

\vspace{3cm}



\newpage
\setcounter{page}{1}

\vspace{0.3cm}
{\large \bf 1. Introduction}

Critical QCD at nonzero baryon density and high temperature is a theory of
fundamental significance associated with the properties of primordial matter
(quark matter) both in the early universe and in the multiparticle
environment, created in heavy-ion collisions. Theoretical arguments, based on
the universal features of chiral phase transition, suggest with confidence
the existence of a critical point in QCD which is the end point of a
quark-hadron critical line of first order and belongs to the Ising
universality class [1]. The location of the critical point in the
phase diagram needs a more detailed treatment, based on first principles and
it is only recently that a first attempt to locate the critical point, within
the framework of lattice QCD, has appeared [2]. It is of interest to note that
the knowledge of the location of the critical point in the phase diagram is
of primary importance for heavy-ion physics since one expects to observe
spectacular critical fluctuations only in those collisions of nuclei which
freeze-out close to the critical point [3].

In this Letter we present a different, complementary approach in the search
for the existence and location of the critical point, based on the statistical
bootstrap principle [4-6]. We notice that in the vicinity of the QCD critical
point (in the crossover regime) the hadronic degrees of freedom are
thermodynamically indistinguishable from the quark matter degrees of freedom,
suggesting that the QCD critical point must also be present in the solution
associated with the statistical bootstrap equations. These equations describe
the dynamics of interacting hadronic matter at high temperatures,
incorporating in the partition function the observed hadronic spectrum [4-6].

In the context of the Statistical Bootstrap the interaction among
hadrons is simulated by the presence of clusters. The bootstrap hypothesis
adopts that every cluster is composed by a number of other clusters described
by the same mass spectrum $\tilde{\tau}(m^2,b,s,\ldots)$, where
$b$ stands for baryon number, $s$ for strangeness and the dots for any
additional quantum number that can characterise a cluster, like
charge $q$, absolute strangeness (number of strange plus antistrange quarks)
$|s|$, etc. Then $\tilde{\tau}dm^2$ equals to the number
of clusters with mass $m$ and quantum numbers equal to the set $b,s,\ldots$.
Every cluster is also characterised by a kinematical term $\tilde{B}(m^2)$,
apart from its mass spectrum which is of dynamical origin. The
conservation of four-momentum, as well as, conservation of quantum numbers
can be accommodated by a suitable bootstrap equation [7]
\[
\tilde{B}(p^2)\tilde{\tau}(p^2,b,s,\ldots)=
\underbrace{g_{b,s,\ldots}\tilde{B}(p^2)\delta_0(p^2-m_{b,s,\ldots}^2)}_
{input\;term}
+\sum_{n=2}^\infty\frac{1}{n!}\int\delta^4\left(p-\sum_{i=1}^np_i\right)
\cdot
\]
\begin{equation} 
\cdot\sum_{\{b_i\}}\delta_K\left(b-\sum_{i=1}^n b_i\right) 
\sum_{\{s_i\}}\delta_K\left(s-\sum_{i=1}^n s_i\right) \ldots 
\prod_{i=1}^n\tilde{B}(p_i^2)\tilde{\tau}(p_i^2,b_i,s_i,\ldots)d^4p_i\;. 
\end{equation} 

The pattern of clusters composed of clusters continues until
the known hadrons, considered as the ``input'' particles, are reached.
The bootstrap equation after performing a series of
Laplace transformations acquires the form\footnote{We shall use the
abbreviation $\{\lambda\}$ for the set of fugacities corresponding to the
quantum numbers $(b,q,s,|s|)$, i.e.
($\lambda_b,\lambda_q,\lambda_S,\gamma_s$).}.
\begin{equation} 
\varphi(\beta,\{\lambda\})=2G(\beta,\{\lambda\}) 
-\exp [G(\beta,\{\lambda\})]+1\;. 
\end{equation}
In the last equation $G(\beta,\{\lambda\})$ is the Laplace transform of the
mass spectrum with the accompanying kinematical term
\[
G(\beta,\{\lambda\})=
\sum_{b'=-\infty}^{\infty}\lambda_b^{b'}
\sum_{q'=-\infty}^{\infty}\lambda_q^{q'}
\sum_{s'=-\infty}^{\infty}\lambda_S^{s'}
\sum_{|s'|=0}^{\infty}\gamma_s^{|s'|}
\int e^{-\beta^\mu p_\mu}\tilde{B}(p^2)\tilde{\tau}(p^2,b',q',s',|s'|)dp^4\;,
\]
\begin{equation} 
\hspace{1cm}=\frac{2\pi}{\beta}\int_0^{\infty} m\tilde{B}(m^2) \tilde{\tau}
(m^2,\{\lambda\})K_1(\beta m)dm^2 
\end{equation} 
and $\varphi(\beta,\{\lambda\})$ the Laplace transform of the input
term\footnote{The Statistical Bootstrap Model is formulated in the
Boltzmann approximation which is acceptable throughout the phase diagram
except for low temperatures and high baryon densities, where the effect of
the Bose-Einstein and Fermi-Dirac statistics becomes important [5].
For example for the temperature and baryon density which correspond to
SPS the error due to the neglection of the correct statistics does
not exceeds $3\%$ [7].}
\begin{equation}
\varphi(\beta,\{\lambda\})=
\sum_{b'=-\infty}^{\infty}\lambda_b^{b'}
\sum_{q'=-\infty}^{\infty}\lambda_q^{q'}
\sum_{s'=-\infty}^{\infty}\lambda_S^{s'}
\sum_{|s'|=0}^{\infty}\gamma_s^{|s'|}
\int e^{-\beta^\mu p_\mu}g_{b'q's'|s'|}\tilde{B}(p^2)\delta_0(p^2-m_{b'q's'|s'|}^2)
dp^4\;. 
\end{equation} 
The masses $m_{bqs|s|}$ correspond to the masses of the input
particles, which in this paper will be all the known
hadrons with masses up to 2400 MeVs, the $g_{bqs|s|}$ are degeneracy factors
due to spin and the $\lambda$'s are the fugacities of the input particles.

The bootstrap equation [7] defines the boundaries of the hadronic phase
through the relation
\begin{equation}
\varphi(\beta,\{\lambda\}) \le \ln4-1\;.
\end{equation} 

The same equation leads to the asymptotic mass spectrum of
interacting hadrons
\begin{equation} 
\tilde{\tau}(m^2,\{\lambda\})\stackrel{m\rightarrow\infty} 
{\longrightarrow} 
2C(\{\lambda\})m^{-\alpha-1} \exp [m\beta^*(\{\lambda\})]\;, 
\end{equation} 
where $\beta=T^{-1}$ and $\beta^*$ corresponds to the maximum inverse
temperature [7].
The kinematical term $\tilde{B}$ and the dynamical term $\tilde{\tau}$
can be redefined so that their product remains unchanged. The different
choices adopted lead to different values of the exponent $\alpha$ in eq. (6)
and to different physical behaviour of the system through different partition
functions. In this Letter we shall fix $\alpha=4$. This choice leads to normal behaviour of the system near the
boundaries of the hadronic phase since it does not allow the energy density
to become infinite, even for pointlike particles [6]. Consequently, for this
choice, the maximum
hadronic temperature is not a limiting temperature, since it can be reached at
finite energy density and so it is consistent with a phase transition [8]. It
also guarantees that
the bootstrap singularity associated with a quark-hadron phase transition is
reached in the thermodynamic limit [9].

\vspace{0.3cm}
{\large \bf 2. The existence of a critical point}

For the particular choice of $\alpha=4$, $\varphi$ assumes the form
\begin{equation} 
\varphi(\beta,\{\lambda\})=\frac{1}{2\pi^2 \beta B} 
\sum_{\rm a} (\lambda_{\rm a}(\{\lambda\})+\lambda_{\rm a}(\{\lambda\})^{-1})
\sum_i g_{{\rm a}i} m_{{\rm a}i}^3 K_1 (\beta m_{{\rm a}i})\;\;, 
\end{equation} 
where ``${\rm a}$'' represents a particular hadronic family and ``$i$'' the
hadrons in the family with different masses.
The partition function of the pointlike interacting hadrons for this choice
of $\alpha$ reads [7]:
\[ 
\ln Z(V,\beta,\{\lambda\})= 
\sum_{b'=-\infty}^{\infty}\lambda_b^{b'}
\sum_{q'=-\infty}^{\infty}\lambda_q^{q'}
\sum_{s'=-\infty}^{\infty}\lambda_S^{s'}
\sum_{|s'|=0}^{\infty}\gamma_s^{|s'|}
\int \frac{2V_{\mu}p^{\mu}}{{(2\pi)}^3}
\tilde{\tau}(p^2,b',q',s',|s'|) e^{-\beta^\mu p_\mu} dp^4= 
\]
\begin{equation}
\frac{4BV}{\beta^3}\int_{\beta}^{\infty} x^3 G(x,\{\lambda\})dx \equiv
Vf(\beta,\{\lambda\})\;,
\end{equation} 
where the only left parameter $B$ is the energy density of the vacuum (bag
constant). Including corrections due to the finite size of hadrons (Van der
Waals volume corrections) one takes into account the repulsive part of the
interaction [10]. With this improvement, the partition function in the grand
canonical pressure ensemble, given by the Laplace transform of
$Z(V,\beta,\{\lambda\})$, is written [10]
\begin{equation} 
\pi(\xi,\beta,\{\lambda\})=
\left[\xi-f\left(\beta+\frac{\xi}{4B},\{\lambda\}\right)\right]^{-1}\;,
\end{equation}
where $f$ is the point-like partition function (8) divided by the volume and
the Laplace variable $\xi$ gives a measure of the external forces
acting on the system. In what follows we consider hadronic systems free of
external forces and therefore we fix the variable $\xi$ at the value $\xi=0$
[9]. From eqs. (8) and (9) one obtains the energy and baryon number density
$(\varepsilon,\nu_b)$ and the pressure $P$ of the thermodynamic system with
the help of the following equations:
\[ 
\varepsilon(\beta,\{\lambda\})=
-\frac{\partial f}{\partial \beta}
\left[1-\frac{1}{4B}\frac{\partial f}{\partial \beta}\right]^{-1}
\]
\begin{equation} 
\nu_b^{(h)}(\beta,\{\lambda\})=
\lambda_b\frac{\partial f}{\partial \lambda_b}
\left[1-\frac{1}{4B}\frac{\partial f}{\partial \beta}\right]^{-1}
\end{equation}
\[ 
P^{(h)}(\beta,\{\lambda\})=
\frac{1}{\beta}f(\beta,\{\lambda\})
\left[1-\frac{1}{4B}\frac{\partial f}{\partial \beta}\right]^{-1}
\]
The onset of a phase transition can be traced by studying the
pressure-volume isotherm curve (Fig. 1). It is obtained with the help of
eqs. (10) and it reveals a region of instability where pressure and volume
decrease together. This is a signal of a first-order phase transition and is
due to the formation of bigger and bigger clusters as the hadronic matter
approaches its boundaries. This instability is removed by a Maxwell
construction as it is shown in Fig. 2 and discussed in more detail in what
follows. As expected, no such behaviour is exhibited if the interaction
incorporated in the statistical bootstrap hypothesis (SB) is neglected. In
fact, in an ideal hadron gas (IHG), even with Van der Waals volume
corrections, the pressure is everywhere a decreasing function of volume,
keeping the system far from a phase transition. The results are displayed in
Fig. 1, where $\nu_0$ is the normal nuclear density $\nu_0=0.14\;fm^{-3}$.
For the ideal hadron gas we have used in place of
$f(\beta,\{\lambda\})$ the function:
\begin{equation} 
\frac{\ln Z_{IHG}(V,\beta,\{\lambda\})}{V}=\frac{1}{2\pi^2\beta} 
\sum_{\rm a} [\lambda_{\rm a}(\{\lambda\})+\lambda_{\rm a}(\{\lambda\})^{-1}]
\sum_i g_{{\rm a}i} m_{{\rm a}i} K_2 (\beta m_{{\rm a}i})\;, 
\end{equation}
where $g_{{\rm a}i}$ are degeneracy factors due to spin and isospin. In both
cases (SB, IHG) we have imposed the constraints: $<S>=0$ (zero strangeness)
and $<b>=2<q>$ (isospin symmetric system, i.e. the net number of $u$ and $d$
quarks are equal).

The existence of a maximal pressure in the isotherm curves (Fig. 1), obtained
in the bootstrap solution (eqs. (10)), is the starting point in the search for
the QCD critical point. In fact the end point of the critical line of a first
order transition is reached when two isotherm curves corresponding to the two
phases and the same temperature, intersect at the point of maximal pressure,
on the hadronic side. The point of intersection, in this case, coincides with
the critical point which in general corresponds to nonzero baryon chemical
potential. In order to quantify the above description we have to consider the
partition function of the quark-gluon system (QGP) and study the
pressure-volume isotherm curves. For this purpose we adopt a minimal
description of non interacting quarks and gluons within a bag supported by an
external pressure $B$. More specifically, the free energy is
written [11]\footnote{The set of quark flavour fugacities
$(\lambda_u,\lambda_d,\lambda_s,\gamma_s)$ can be used equivalently in the
place of the set $(\lambda_b,\lambda_q,\lambda_S,\gamma_s)$ [7].}
\begin{eqnarray}
\beta^{-1}\ln Z_{QGP}(V,\beta,\lambda_u,\lambda_d)=&&\hspace{-0.5cm}
\frac{N_s N_c V}{6\pi^2}
\sum_i \int_0^{\infty} \frac{p^4}{\sqrt{p^2+m_i^2}}
\frac{1}{e^{\beta \sqrt{p^2+m_i^2}}\lambda_i^{-1}+1}dp \nonumber \\
&&\hspace{-0.5cm}+V\frac{8\pi^2}{45}\beta^{-4}-BV\;.
\end{eqnarray}
where the index $i$ runs over all quarks and antiquarks and $N_s$, $N_c$
refer to spin and colour degrees of freedom. The current masses are taken
$m_u\approx 5.6$ MeV, $m_d\approx 10$ MeV, $m_s\approx 200$ MeV and the
fugacities satisfy the constraints: $\lambda_u=\lambda_{\bar{u}}^{-1}$,
$\lambda_d=\lambda_{\bar{d}}^{-1}$, $\lambda_s=\lambda_{\bar{s}}^{-1}=1$
(strangeness is set to zero). The first two terms in eq. (12) give the
contribution to the free energy of the noninteracting quarks and gluons
respectively and the third term gives the contribution of the vacuum (bag
constant).

In order to study the $P-V$ isotherms of the quark-gluon
system, we extract from the free energy (12) the baryon number density in
quark matter, as follows
\begin{equation} 
\nu_b^{(q)}(\beta,\lambda_u,\lambda_d)=
\frac{N_s N_c}{2\pi^2} \sum_i \lambda_i N_i
\int_0^{\infty}\frac{p^2 dp}{e^{\beta\varepsilon_i}+\lambda_i}\;.
\end{equation}
The index $i$ includes only $u$, $\bar{u}$ and $d$, $\bar{d}$ quarks whereas
$N_i=\pm 1$ for quarks $(u,d)$ and antiquarks $(\bar{u},\bar{d})$
correspondingly. For a given value of $B$, the point where the hadron and
quark phases meet can be traced on the $P-V$ isotherms for different
temperatures. One has to require that at the same baryon density
$\nu_b^{(q)}=\nu_b^{(h)}$ (eqs. (10) and (13)) the pressures, for the two
phases, coincide. As can be seen in Fig. 2, for a low temperature, the quark
matter isotherm meets the corresponding hadron isotherm at a point where the
pressure is an increasing function of volume. This region is associated with
the onset of a first-order phase transition and a Maxwell construction is
needed in order to remove the instability [6,9]. As the temperature increases
the quark matter isotherm meets the hadronic isotherm at a point which
corresponds to a higher density and pressure. With this set of intersection
points, the critical line of a first order quark-hadron phase transition is
approximately built up, in the phase diagram. At a certain temperature, the
two phases meet at the point of maximal hadronic pressure (Fig. 2). The
Maxwell construction is no longer needed since we have reached the end point
of the critical line, in other words, we have reached a critical point of the
strongly interacting matter. For higher temperatures we enter the crossover
regime where quark and hadron phases are no longer distinguishable.

In Fig. 2 the limitations of our approach are also illustrated. In fact, in
the absence of a final theory, based on QCD, which could provide us with a
unified treatment of the equation of state in both phases, the matching of
the isotherms, shown in Fig. 2, introduces artificial discontinuities in the
derivatives for $T>T_c$, in a region where singularities are not allowed
(crossover regime). Moreover, the scaling behaviour expected at the critical
point and expressed through the isotherm critical exponent ($\delta$), is not
recovered at the point of intersection for $T=T_c$ (Fig. 2). Despite these
drawbacks, our treatment concerning the existence and location of the
critical point may still remain valid if in the exact theory of the equation
of state, not only the unwanted discontinuities are washed out but, at the
same time, the qualitative features of the hadronic isotherm, predicted by
Statistical Bootstrap (Fig. 1), remain unchanged.

\vspace{0.3cm}
{\large \bf 3. The location of the critical point}

On the basis of the above mechanism and in order to locate the critical point
in the phase diagram $\mu_b - T$, one has to determine the parameters
$(\beta,\lambda_u,\lambda_d,\lambda_s,\lambda_u',\lambda_d')$ by solving the
set of equations:
\[
\hspace{5.5cm}
\nu_b^{(h)}(\beta,\lambda_u,\lambda_d,\lambda_s)=
\nu_b^{(q)}(\beta,\lambda_u',\lambda_d')
\hspace{4cm}(14{\rm a})
\]
\[
\hspace{5.4cm}
P^{(h)}(\beta,\lambda_u,\lambda_d,\lambda_s)=
P^{(q)}(\beta,\lambda_u',\lambda_d')
\hspace{4cm}(14{\rm b})
\]
\[
\hspace{6.4cm}
\frac{\partial P^{(h)}(\beta,\lambda_u,\lambda_d,\lambda_s)}
{\partial \lambda_u}=0
\hspace{4.9cm}(14{\rm c})
\]
\[
\hspace{6.4cm}
\left<\;S(\beta,\lambda_u,\lambda_d,\lambda_s)\;\right>_h=0
\hspace{4.9cm}(14{\rm d})
\]
\[
\hspace{4.4cm}
\left<\;b(\beta,\lambda_u,\lambda_d,\lambda_s)\;\right>_h-
2\;\left<\;Q(\beta,\lambda_u,\lambda_d,\lambda_s)\;\right>_h=0
\hspace{2.9cm}(14{\rm e})
\]
\[
\hspace{4.9cm}
\left<\;b(\beta,\lambda_u',\lambda_d')\;\right>_q-
2\;\left<\;Q(\beta,\lambda_u',\lambda_d')\;\right>_q=0
\hspace{3.4cm}(14{\rm f})
\]
Equations (14d)-(14f) account for isospin symmetry and zero strangeness.

For a given value of the bag constant $B$ we have investigated the solution
of eqs. (14) in order to build up the critical line (excluding eq. 14c) and
locate the endpoint (including eq. 14c) in the phase diagram. We have
projected out the solution onto the plane $\mu_b-T$ and our first remark is
that for a wide range of values of the bag constant $B$ ($B^{1/4}<282$ MeV)
the critical baryon chemical potential $\mu_c$ is not zero (Fig. 3a). One may
therefore establish, on the basis of the statistical bootstrap hypothesis,
the existence of a critical point in hadronic matter at high temperatures and
nonzero baryon chemical potential. A typical solution is obtained if we
impose the condition $\mu_b$(hadronic matter) = $\mu_b$(quark matter) at the
critical point. The solution
obtained in this case corresponds to the value $B^{1/4}=250$ MeV and it is
illustrated (SB) in Fig. 3b. The location of the endpoint (critical point) in
this solution is fixed by the critical values $T_c \approx 171 MeV$ and
$\mu_c \approx 385$ MeV whereas the critical temperature at zero chemical
potential (at the end of the crossover) is $T_0 \approx 188$ MeV. In the same
figure, the solution of lattice QCD with $n_f=2+1$ (LQCD), corresponding to
the critical values $T_c = 160 \pm 3.5$ MeV, $\mu_c = 725 \pm 35$ MeV and
$T_0=172$ MeV is also shown [2] for comparison. The discrepancy of these
two solutions is due (a) to the fact that the lattice QCD result for the
critical point is based on rather unphysical input values of the quark masses
$m_u$, $m_d$ and (b) to the uncertainties in the partition function (12)
coming from neglected interactions in the quark-gluon system. One expects
that, when both, the quark masses get reduced and the quark-gluon pressure for
nonzero chemical potential becomes available on
the lattice [12], the SB and LQCD critical lines in Fig. 3b will converge to a
unique solution, supported both by QCD and Statistical Bootstrap. Finally, it
is of interest to note that within a different frame of approximate theories
(Nambu-Jona-Lasinio (NJL) model or a random matrix (RM) approach) the QCD
critical point has been located in the region $T_c \approx 100$ MeV,
$\mu_c \approx 600-700$ MeV of the phase diagram [1], in a good distance from
the SB and LQCD solutions (Fig. 3b).

In Fig. 3b the freeze-out points associated with recent heavy-ion experiments
are also shown on the plane $\mu_b - T$ [13]. One observes that the SB
solution brings the critical point close to the freeze-out area of the SPS
experiments [13]. However, before drawing any conclusion concerning the
phenomenology of the SB critical point, an improved lattice QCD calculation
of the pressure of the quark-gluon system with small quark masses is needed
[12].

In conclusion, we have shown that the hadronic matter develops a critical
point at high temperatures and nonzero baryon chemical potential, in
accordance with the requirements of the QCD phase diagram. Our approach was
based on the Hagedorn Statistical Bootstrap Principle (SB) as it was
elaborated, in connection with quark-hadron phase transition, in references
[6,9]. With the present degree of approximation for the quark-gluon partition
function, the location of the SB critical point, in the phase diagram,
was found close to the freeze-out points of the SPS experiments with nuclei
($T_c \approx 171$ MeV, $\mu_c \approx 385$ MeV).

{\large \bf Acknowledgements}

It is a pleasure to thank Fotis Diakonos for fruitful discussions. This work
was supported in part by the Research Committee of the University of Athens.

{\bf Figure Captions}
\newtheorem{f}{Fig.} 
\begin{f} 
\rm The pressure-volume isotherms for hadronic matter described by
    Statistical Bootstrap (SB) and for the ideal hadron gas (IHG).
\end{f} 
\begin{f} 
\rm The intersection points of the pressure-volume isotherms of the two
    phases are shown, for different temperatures, and the Maxwell
    construction associated with the phase transition is illustrated.
\end{f} 
\begin{f} 
\rm (a) The SB solutions for the critical line and the corresponding critical
    point, for different choices of the bag constant.
    (b) A typical SB solution is shown in the phase diagram ($\mu_b-T$)
    together with the LQCD solution and the freeze-out points of heavy-ion
    experiments. Also the NJL-RM solution for the location of the critical
    point is shown.
\end{f}

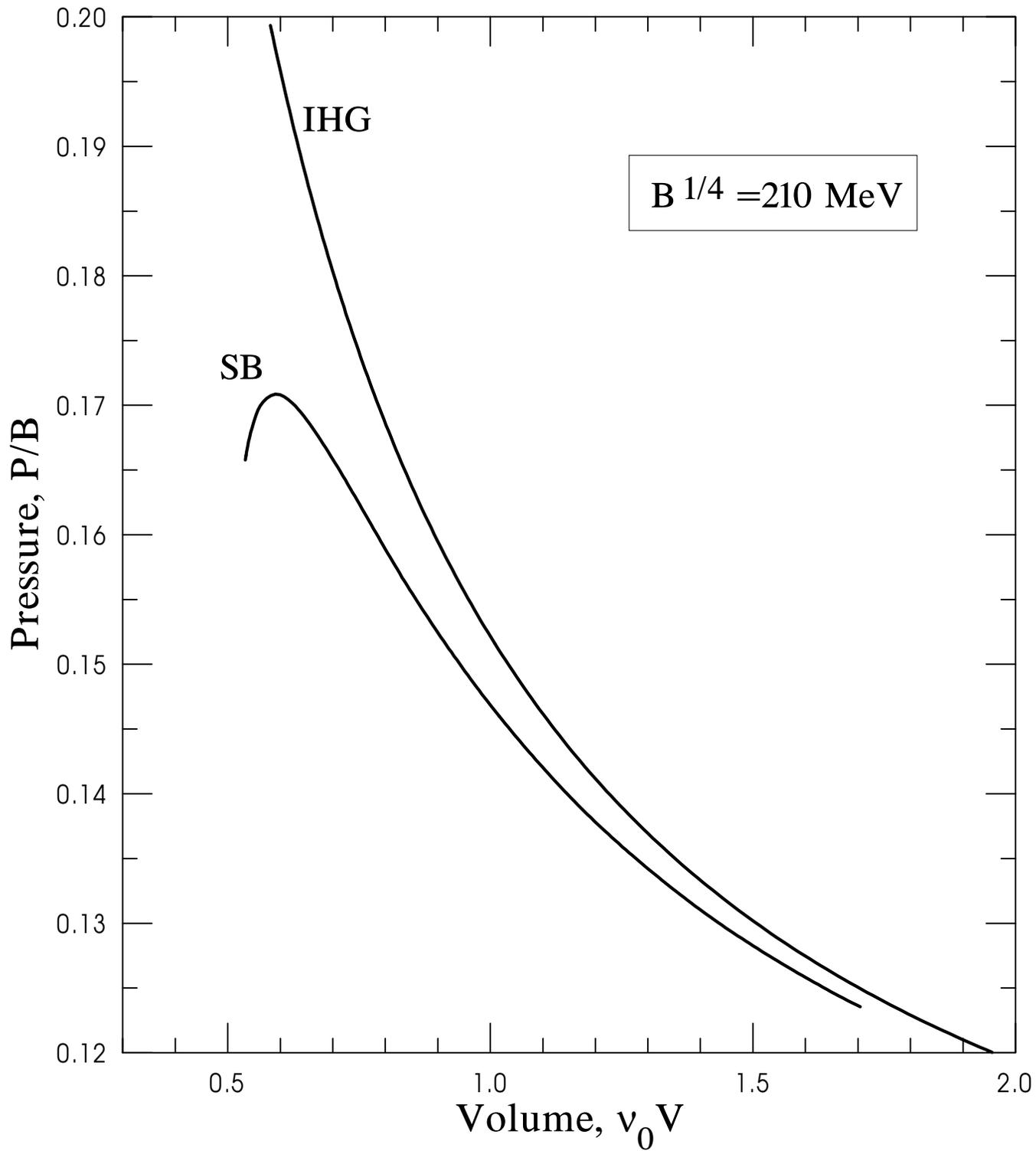
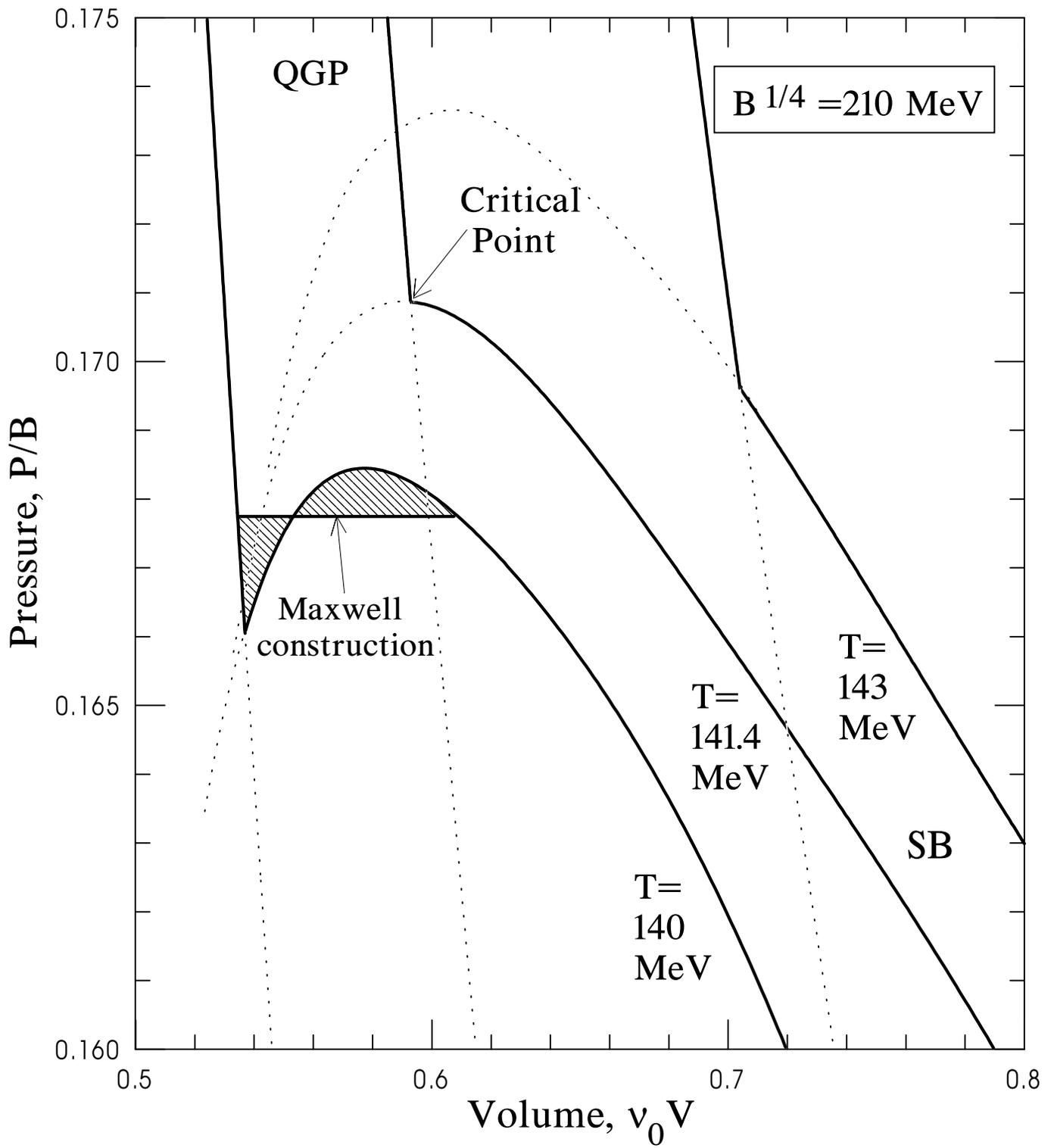
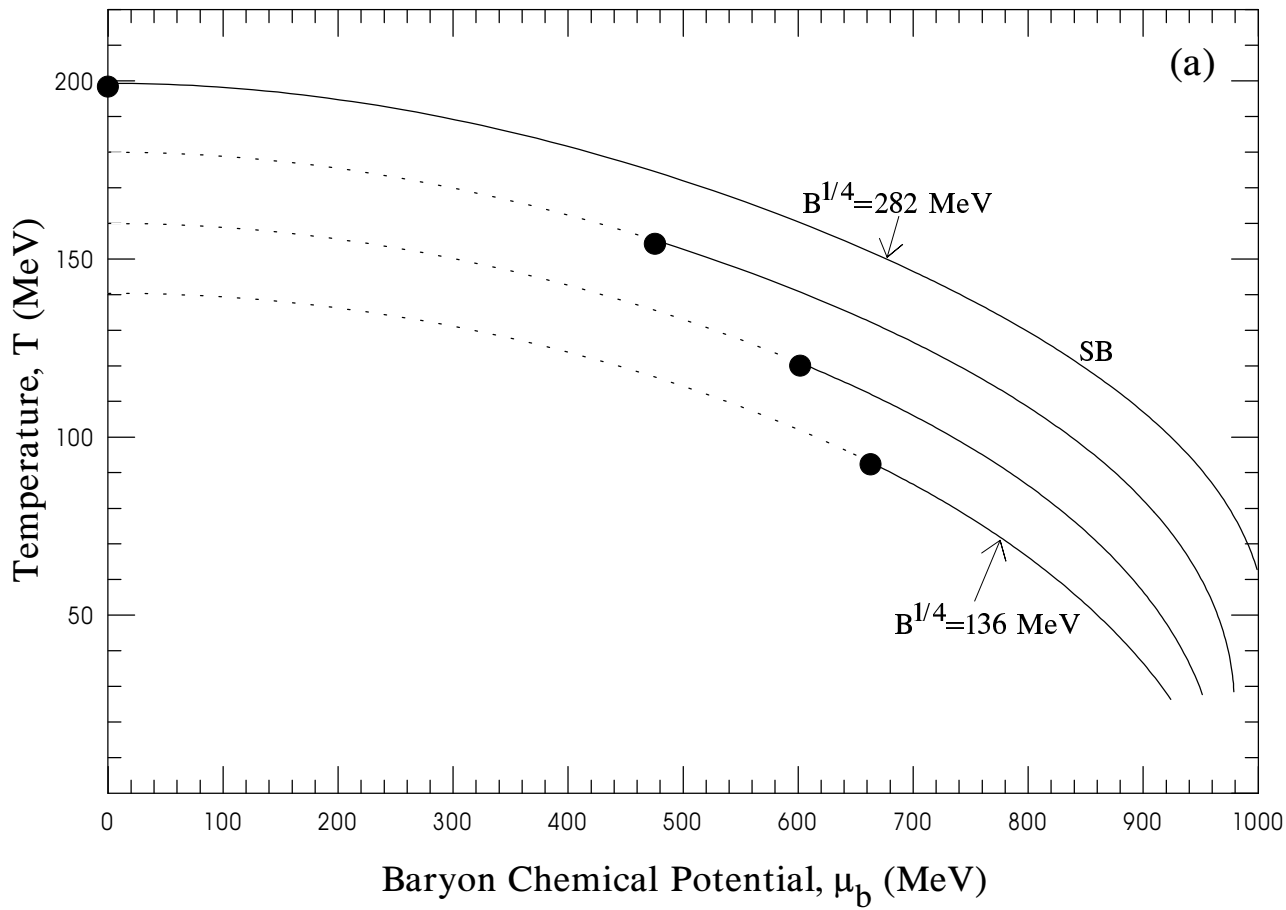
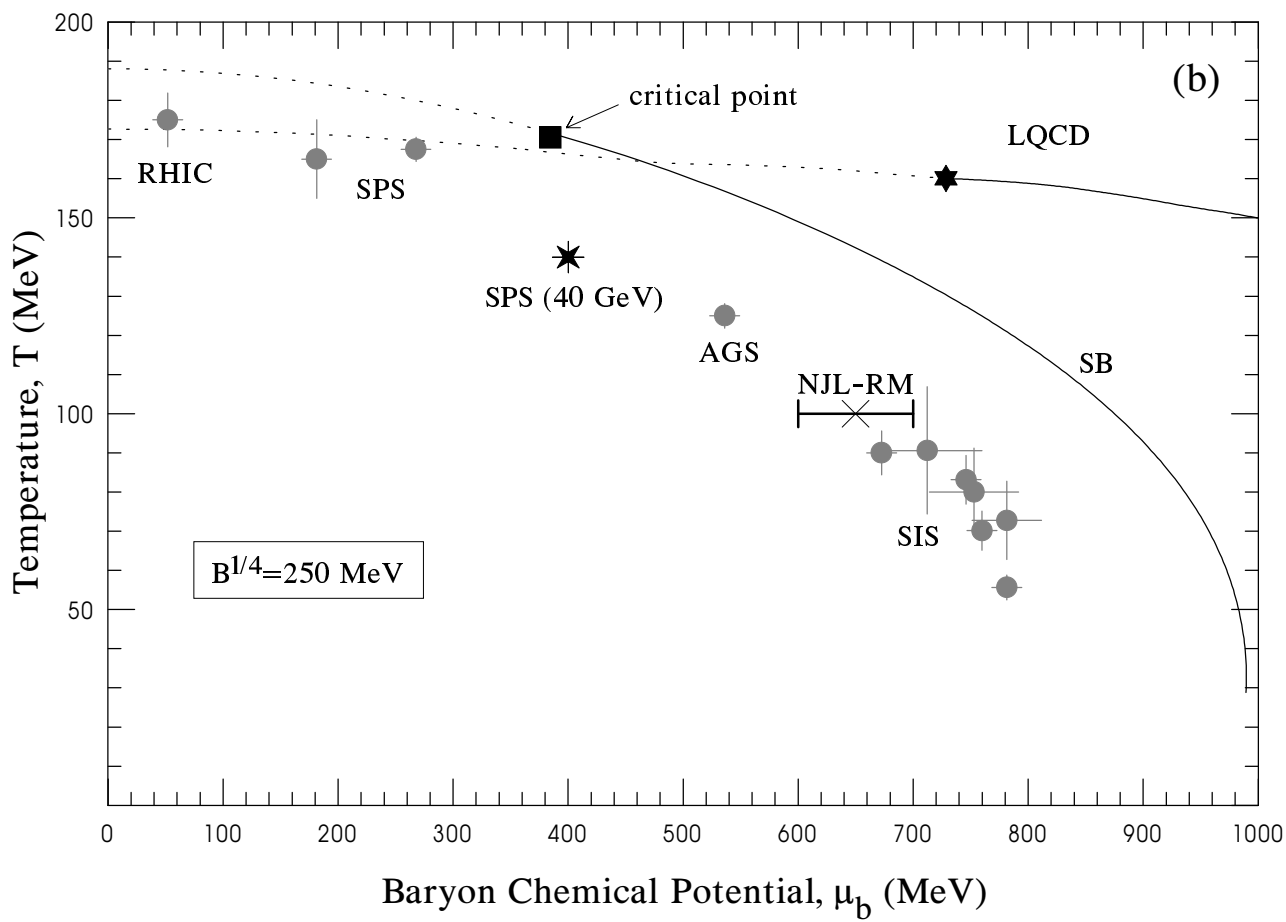
\end{document}